\documentclass[11pt,twoside]{article}
\usepackage{asp2010}

\resetcounters

\usepackage{graphicx}
\usepackage[breaklinks]{hyperref}
\usepackage{natbib}  

\begin{document}

\allowtitlefootnote

\title{The Formation of the Milky Way Nuclear Cluster}
\author{A.~Mastrobuono-Battisti, R.~Capuzzo-Dolcetta }
\affil{Dep. of Physics, Sapienza, Universit\`{a} di Roma, Italy\\
}

\begin{abstract} 
%
Nuclear Star Clusters are observed at the center of many galaxies.
In particular, in the Milky Way's center, the Nuclear Star Cluster coexists with a central supermassive black hole. 
The origin of these clusters is still unknown; a possible formation mechanism is the decay of massive globular clusters driven inward to the galactic center by dynamical friction and their subsequent merging.
By investigating this scenario  by means of sophisticated N-body simulations we found that this process could lead to a final product which actually shows many of the observed features of the Milky Way Nuclear Star Cluster.

\end{abstract}
%
%
%
\section{Introduction} 
Nuclear Star Clusters (NSCs) have been observed in many early-type galaxies and late-type spirals. 
The Milky Way (MW) and a handful of other galaxies contain both a NSC and a massive or even super massive black hole (MBH).
NSCs are among the densest star clusters observed, with effective radii of a few pc and central luminosities up to $\sim 10^7$~{\rm L}$_\odot$.
The formation mechanism of NSCs is still unknown.
Two competing models are possible.
In the gas model a NSC can form from the gas that
migrates to the center of the galaxy, where  then forms stars \citep{Sc08}.
Due to the inherent complexity of gas dynamics, the
gas  model is almost qualitative and difficult to study.
Alternatively, in the merger model,  massive clusters decay to the center via dynamical friction and merge to form a dense nucleus \citep{tre75, capdol93}.  Observations of NSCs in dE galaxies
suggest that the majority, but not all, dE nuclei could be the result  of packing mass in form of orbitally decayed globular clusters  (GCs)  \citep{lot04}.  Numerical simulations have also shown that the basic properties of  NSCs, including their shape, mass density profile, and mass-radius relation, 
 are well reproduced in the merger model under a variety of  explored conditions \citep{capdol08b,capdol08a,Har11}.

\section{The Galactic NSC: formation and evolution}
The MW  NSC has an estimated mass of $\sim 10^{7}~{\rm M}_{\odot}$ (\citealt{LauZy}), and it hosts an MBH whose mass, $\sim 4.3\times 10^6~{\rm M}_\odot$, is uniquely well determined \citet{Gil}. 
Since the relaxation time at SgrA*'s influence radius is robustly estimated to be $20-30$ Gyr \citep{mer10}, the NSC has not had enough time to settle on the \citet{1976ApJ...209..214B} (BW) cusp stable state.
Indeed, the late-type stellar population exhibits not a cusp 
but, rather, a space density which is flat or even falling toward SgrA*, 
inside a core of radius $0.5~$pc \citep{buc09}. 
\subsection{The simulations}
For the first time, we performed self-consistent  $N$-body simulations to test the merger model for the formation of the MW NSC.
These simulations, unlike in previous works, involve the presence of an MBH at the center of the Galaxy. 
\begin{figure}
\center
\includegraphics[scale=0.31]{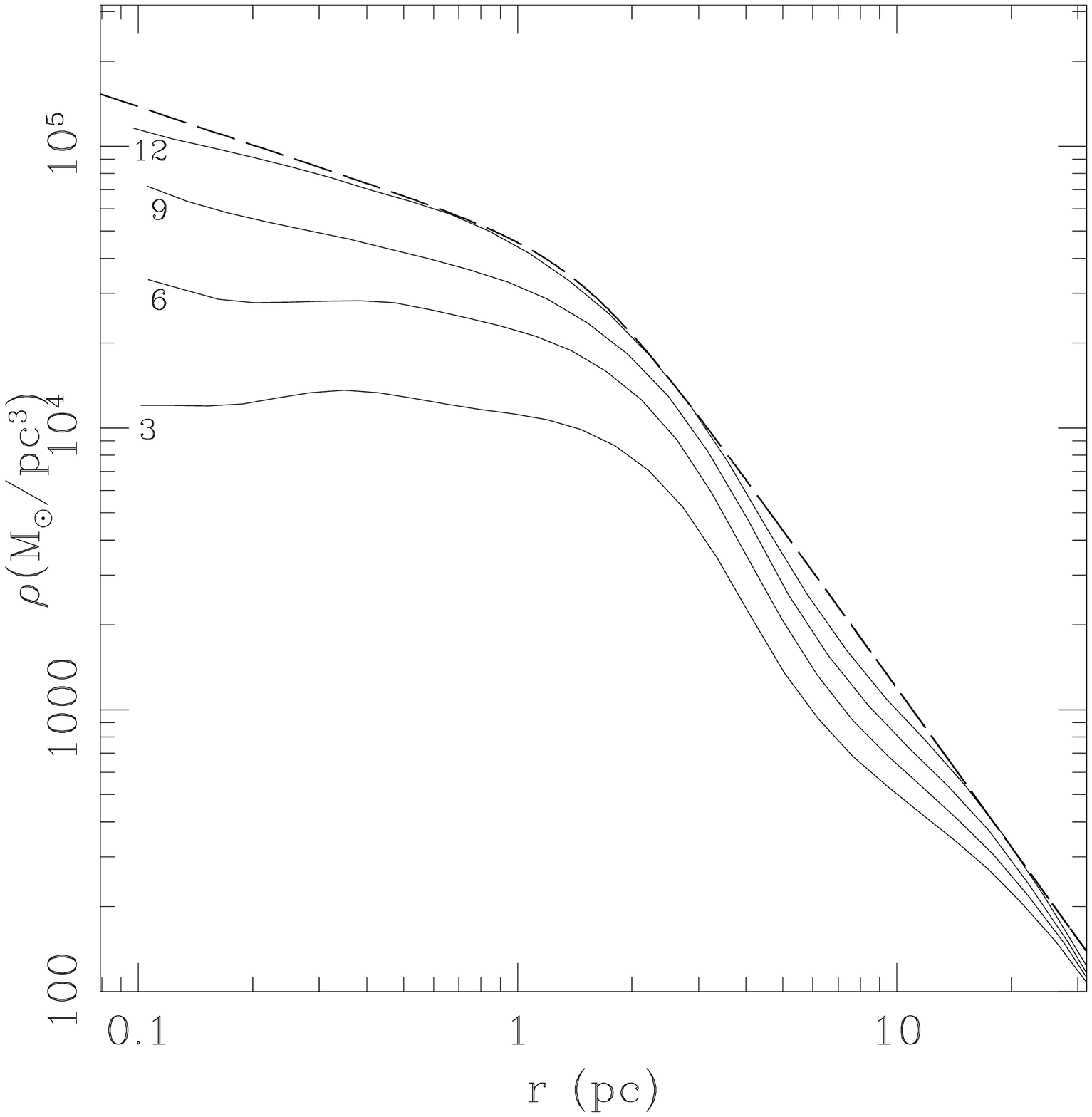} ~
\includegraphics[scale=0.31]{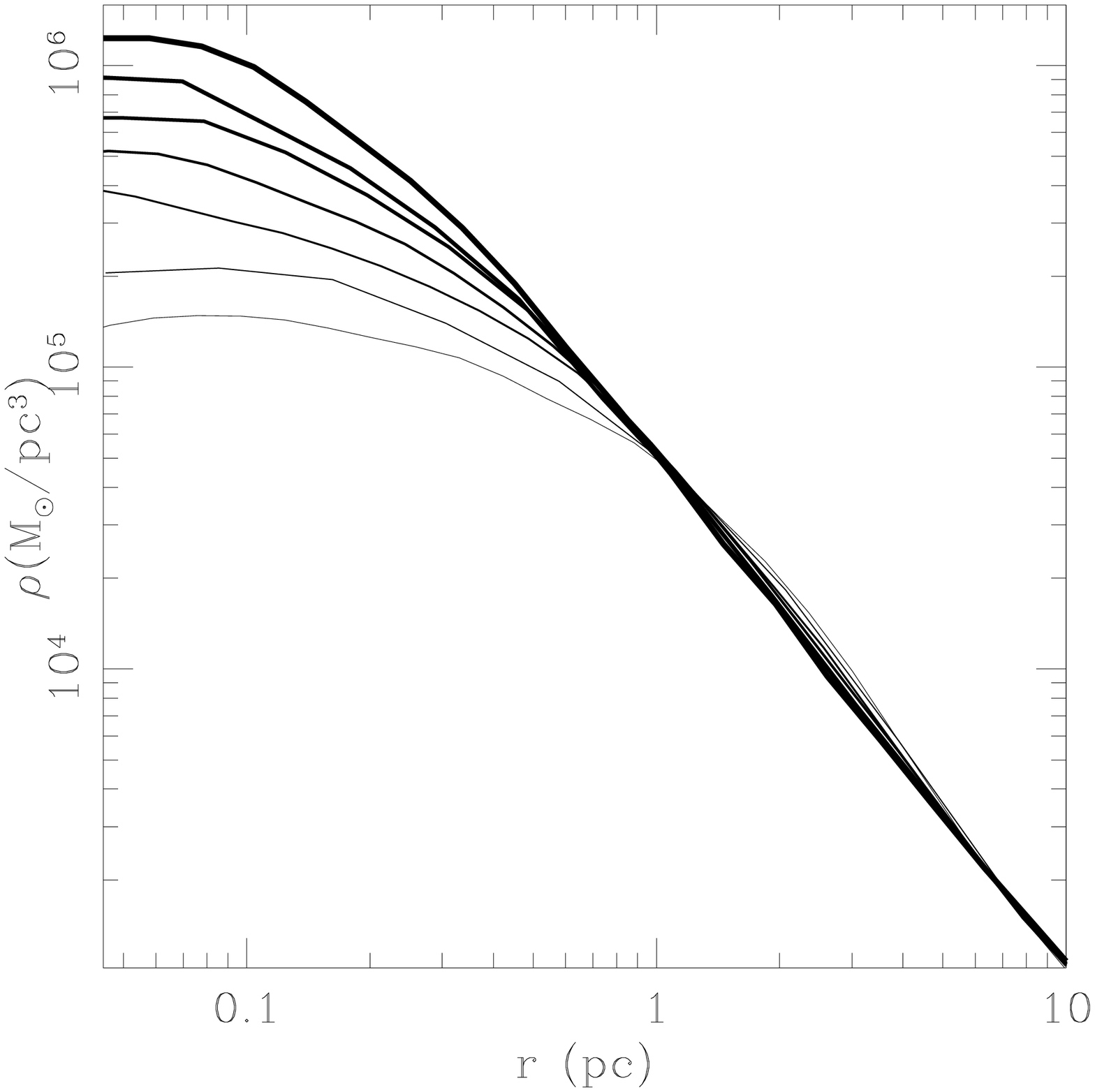}
\caption{\label{fig1} Left panel: Spatial profile of the central NSC after 3, 6, 9 and 12 mergers. The dashed line is the fit to the NSC profile obtained at the end of the last merging event using a broken power law model. Right panel: Spatial profile of the central galactic region at regular intervals of time after the final merging. Time grows upward in the figure.}
\end{figure}
The details of these simulations are described in \citeauthor{ant11} \citetext{2011} and the most significant results are summarized in \citet{CDAM}.
In the simulations, the central MBH is represented as a point mass particle initially located at the origin of coordinates.
The N-body galactic environment is sampled from a shallow density cusp ($\rho\propto r^{-1/2}$), while the individual GC N-body representation has been done starting from a tidally limited, massive and compact King model.
Thus, we performed full N-body simulations to study the consecutive infall and merging of a set of 12 GCs each starting from a galactocentric distance of $20$~pc on an initially circular orbit. This small initial galactocentric distance is the result of previous GC orbital decay due to dynamical friction on a larger spatial scale. After the first GC decayed to the galactic centre, we waited for the NSC to reach a quasi-steady state and we added to the system a second GC. This procedure is iterated until 12 clusters merged in the inner regions of the galaxy.
Thus, the NSC is built up through consecutive infalls of GCs and, after its formation, we followed the long-term evolution of the NSC.
Simulations are done with both the high precision, parallel, NBSymple \citep{cap11} and PhiGRAPE \citep{har07} codes running on the CPU+NVIDIA TESLA C2050 platform in Roma, Sapienza (Italy), and on the Rochester Inst. of Technology (USA) GRAPE cluster. 
\begin{figure}
\center
\includegraphics[scale=0.31]{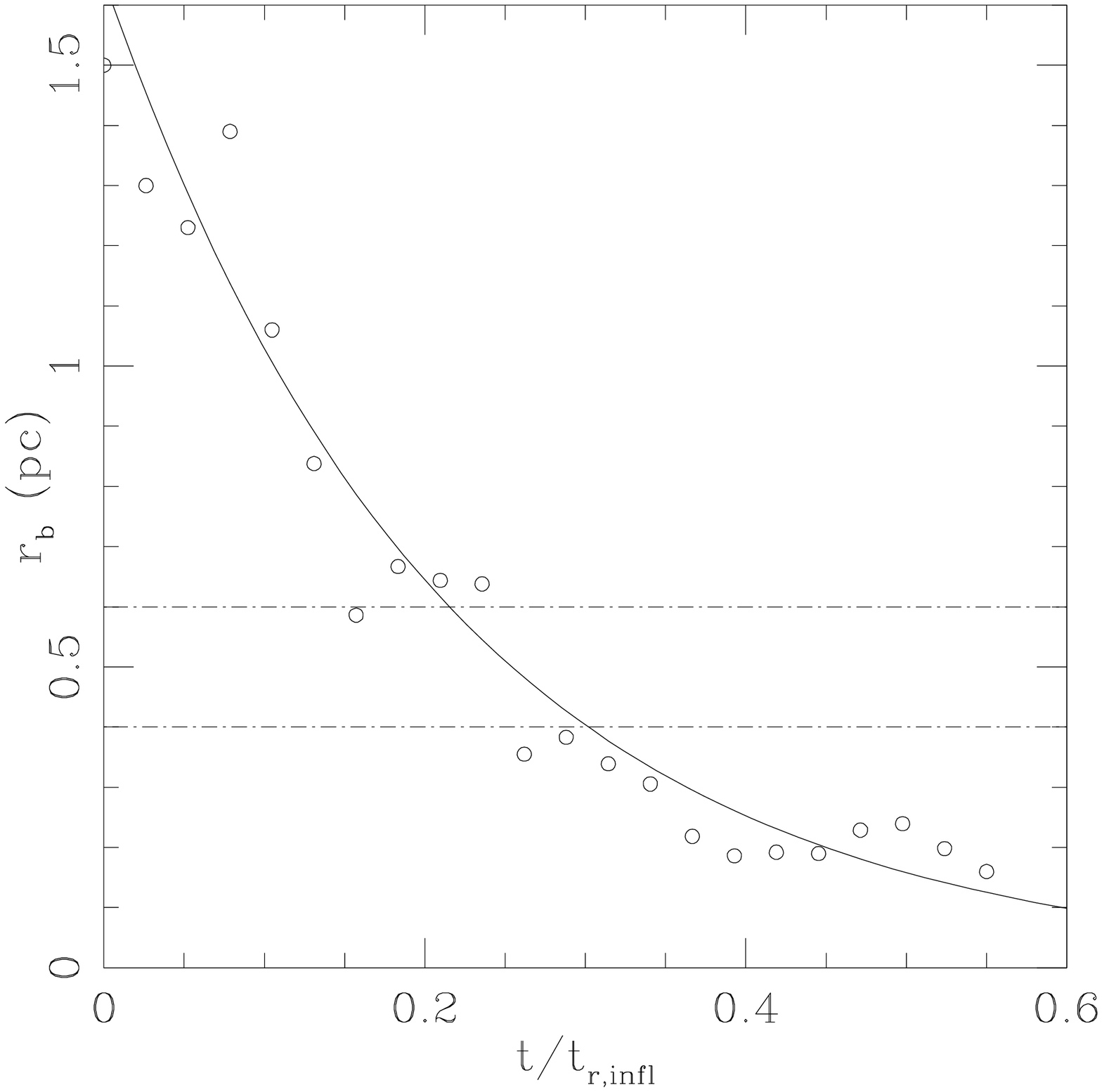} ~
\includegraphics[scale=0.31]{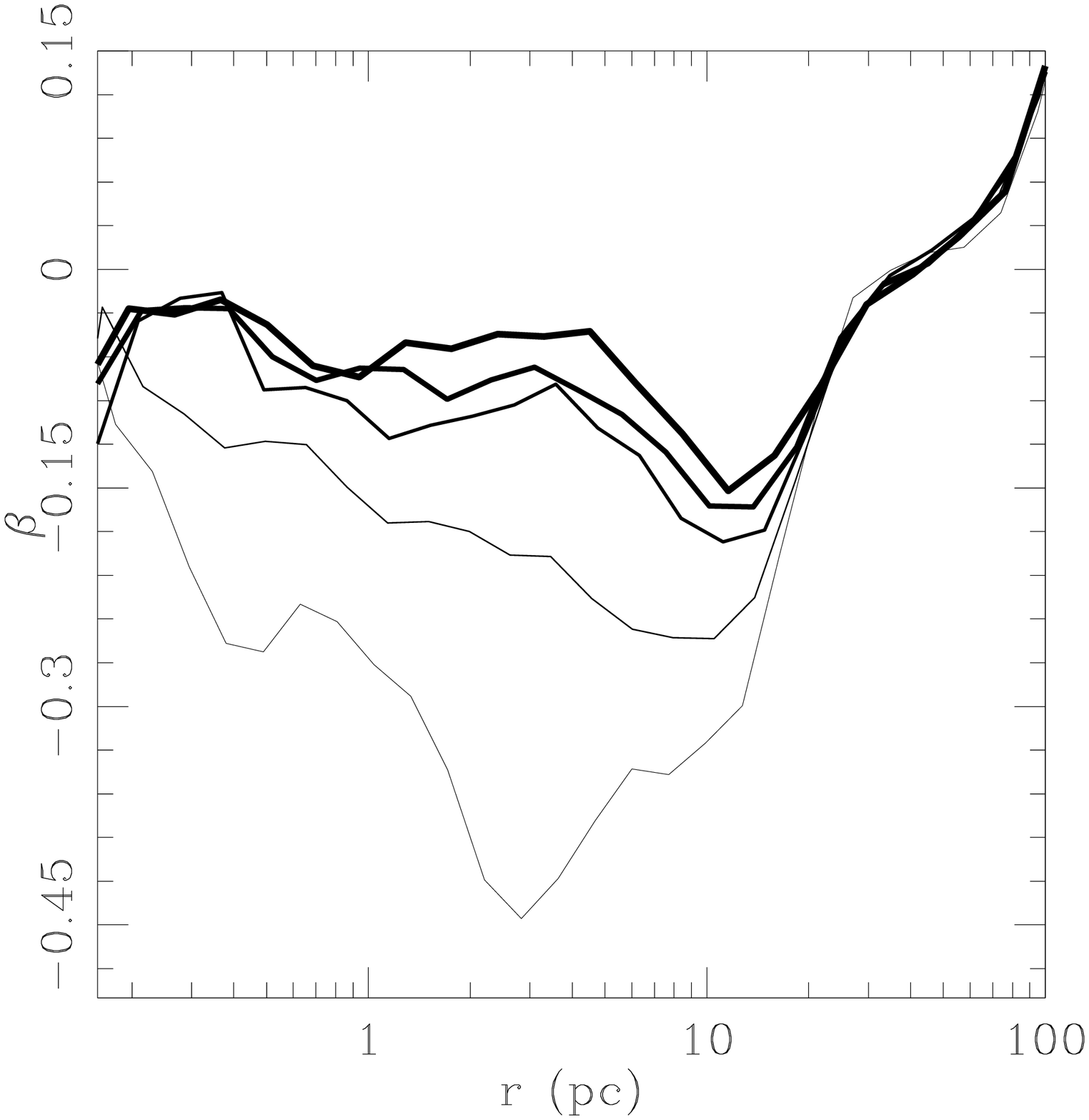}
\caption{\label{fig2}  Left panel: Break radius as a function of time (in units of the $t_{\rm{r,infl}}$) in the evolving model (empty circles). The solid line represents the best exponential fit to the data. The region limited by the two dot-dashed lines corresponds to the two estimates of the MW NSC core radius given in \citet{mer10}. Right panel: Radial profiles of the anisotropy parameter $\beta \equiv 1-(\sigma_\theta^2+\sigma_\phi^2)/(2\sigma_r^2)$ during relaxation displayed at regular interval of times. Thicker curves correspond to older ages.}
\end{figure}
\section{Results}
The merging occurs rather quickly: after about $20$ crossing times the resulting system attains a quasi equilibrium configuration. The orbital energy loss leads to a full merge of the all set of decaying satellites which settle on a meta-stable state, whose following evolution is mainly due to collisional relaxation. 
The left panel of Figure~\ref{fig1} shows the spatial profile of the system after the complete merging of 3, 6, 9 and 12 clusters. As inferred from the best broken power law fit (dashed line in the left panel of Figure~\ref{fig1}), the merger 
product shows a density profile with a central nearly flat core and a slope out of the core region, $\rho \propto r^{-1.8}$, in good agreement with what is observed in the MW \citep{Haller}.  On the other hand, the core radius of the merger remnant is $4$ times larger than the observed value ($\sim 0.5$ pc, \citealt{buc09}).
Anyway we expect that, 
as a natural product of two-body relaxation, such initial parsec-scale core shrinks with time approaching the BW steady state in a time which is of the order of the relaxation time at the influence radius ($r_{\rm infl}$) of the central MBH, $t_{\rm r,infl}$  \citep{mer10}.
Thus,  we followed numerically the collisional relaxation process of the system and found that it undergoes a self-similar evolution which keeps almost unaltered the external radial slope ($\sim-1.8$) while the core shrinks reducing its size (see right panel of Figure \ref{fig1}).
Actually, the evolution of the inner density plateau size of the NSC is shown by the left panel of Figure \ref{fig2}, which gives the break
radius, $r_{b}$, of the best fitting broken power law profile as a function of time. 
The shrinking time dependence of the break (i.e. core) radius is well described by the exponential law
$r_{b}(t)=1.57\,{e}^{-2.24\frac{t}{t_{{\rm r,infl}}}}$.
Right panel of Fig.~\ref{fig2} illustrates the NSC evolution toward isotropy by plotting
the model anisotropy profile at different times.
At the end of the simulation the
NSC has only a small bias toward tangential motion.
This is  consistent  with  proper-motion  data that  indicate a slight degree of tangential anisotropy \citep{S09, mer10}.

\section{Conclusions}
Various authors investigated the merger model for the formation of Nuclear Star Clusters (see for example \citealt{tre75, capdol08b, capdol08a}) but in all these works the presence of a central MBH was neglected and the simulations were not finalized to a specific study of the MW.
Our work covers this  lack through self-consistent simulations whose initial conditions are set up basing on 
recent data about the MW \citep{LauZy}, including the presence of its central MBH.
The merging of GCs produces a system which rapidly settles on a quasi-steady state, slowly evolving due to internal relaxation which, in its turn, is affected by the presence of the MBH.
The central and almost flat core characterizing the density profile of the merger product is maintained for a time 
long enough to justify the core actually observed in the MW NSC. Our results are also supported by the fact that the external slope of the density profile of the NSC remains quite unaltered after the end of the last merger event and its value, $\sim-1.8$, is in optimal agreement with the one inferred from the observations of the Galactic NSC. Moreover, we observe a drift toward velocity isotropy although keeping a slight degree of tangential anisotropy; this peculiarity was found in the observational data and already seen in the numerical study by 
\citet{2001JCoPh.174..208D}.
As a conclusion, the process of consecutive infalls and merging in the inner region of the MW of a number of the order of 10 massive globular clusters could give an explanation of the observed features of the NSC observed in our Galaxy around the Sgr A$^*$ massive black hole.


\bibliography{Mastrobuonobattisti_A}

\clearpage
\end{document}